\documentclass[%
 reprint,
superscriptaddress,
 amsmath,amssymb,
 aps,
]{revtex4-2}
\usepackage{siunitx}
\usepackage{graphicx}  
\usepackage{subfigure}
\usepackage{multirow}
\usepackage{tabularx}
\usepackage{fancyhdr}
\usepackage{longtable}
\usepackage{parskip}
\usepackage[T1]{fontenc}
\usepackage{dcolumn}   
\usepackage{gensymb}
\usepackage{bm}        
\usepackage{amsfonts}  
\usepackage{amsmath}   
\usepackage{amssymb}   
\usepackage{hhline}
\usepackage{xcolor}
\usepackage{ulem}
\usepackage{gensymb}
\usepackage{soul}
\begin{document}

\preprint{APS/123-QED}

\title{Fully Gapped Superconductivity in the Nonsymmorphic Centrosymmetric Superconductor In$_2$Bi\\}

\author{Atul Gangwar}
\affiliation {\it Department of Physics, Indian Institute of Technology (BHU), Varanasi 221005, India}

\author{Kunal Yadav}
\affiliation {\it School of Physical Sciences, Jawaharlal Nehru University, New Delhi 110067, India
}

\author{Barnabha Bandyopadhyay}
\affiliation {\it Department of Physics, Indian Institute of Technology Bombay,
Powai, Mumbai 400076, Maharashtra, India}

\author{S.Patnaik}
\affiliation {\it School of Physical Sciences, Jawaharlal Nehru University, New Delhi 110067, India
}
\author{K.G.Suresh}
\affiliation {\it Department of Physics, Indian Institute of Technology Bombay,
Powai, Mumbai 400076, Maharashtra, India}

\author{Sandip Chatterjee}
\email{schatterji.app@iitbhu.ac.in}
\affiliation {\it Department of Physics, Indian Institute of Technology (BHU), Varanasi 221005, India}


\begin{abstract}
 The nature of superconductivity in nonsymmorphic In$_2$Bi remains an open question, with ongoing debate over whether it exhibits conventional (BCS-like) or unconventional superconductivity.
 Here we report the physical characteristics and superconducting properties of polycrystalline In$_2$Bi. Measurements of electrical resistivity, magnetization, penetration depth, and specific heat demonstrate the superconducting transition temperature, T$_c$ $\approx$ 6 K, with a sharp transition in resistivity ($\Delta T_c=0.1$ K) and RRR = 57.32, reflecting the high quality of the sample. The upper critical field H$_{c2}(0)$ obtained from resistivity and magnetization measurements are in good agreement with each other and are smaller than the Pauli limit, suggesting orbital limited mechanism. The coherence length, $\xi_{GL}(0)=35.43$ nm, penetration depth, $\lambda_{GL}(0)=101.9$ nm and electron-phonon coupling constant, $\lambda_{e-ph}=1.05$ suggest strongly coupled type-II superconductivity. For $T\textless T_c$, $\Delta \lambda(T)$ is best described by an s-wave BCS model and does not have any power law dependency with T, which is expected for a nodeless superconducting order parameter. The superfluid density $\rho_s(T)$ reveals a fully gapped superconductivity with a gap $2\Delta_0/k_B T_c = 4.02$. In the superconducting state, the electronic specific heat is best described by an s-wave BCS model with a gap $2\Delta_0/k_B T_c = 4.05$. All these results from the present study indicate that In$_2$Bi is a typical type-II,BCS-like, strongly coupled, and fully gapped superconductor in the dirty limit.      
\end{abstract}

\maketitle


\section{\label{sec:level1}	INTRODUCTION }

In recent years, a limited number of nonsymmorphic superconductors such as, A15 compounds \cite{Ti3SbRPS,M.KimA15,Ti3SbRchap,A15NbTaTheory}, rare-earth transition-metal ternary borides XRuB$_2$ (X=Y,Lu) \cite{RRuB2,YLuRuB2, YRuB2} and nodal line compounds Yb/CaSb$_2$ \cite{CaSb2nonsymm,YbSb2rps} have emerged as a compelling platform for investigating unconventional pairing mechanisms and topological phenomena arising from crystal symmetry. In conventional centrosymmetric superconductors, the presence of inversion symmetry allows Cooper-pair states to be classified according to their parity, leading, in the simplest case, to a distinction between even-parity spin-singlet and odd-parity spin-triplet pairing. Several type-I superconductors with both centrosymmetric and noncentrosymmetric crystal structures, including AuBe \cite{AuBe1,AuBe2} and LaXSi$_3$ (X=Ir,Rh,Pd,Pt) have consistently revealed conventional, time-reversal-symmetry(TRS)-preserving singlet pairing. Nonsymmorphic symmetries introduce additional spatial constraints through symmetry operations such as screw rotations and glide reflections and protect unconventional electronic band structures, including Dirac crossings and nodal lines, and can profoundly influence the superconducting pairing landscape and its topological properties \cite{PdTecsyadav}. In the presence of strong spin–orbit coupling (SOC), these symmetry elements can generate nontrivial band topology and protected surface states which is the key point to study topological superconductors (TSCs) \cite{RevModPhys.62.1027,RevModTITSC}. It is well known that SOC often arises from the presence of heavy elements and Bi is one of the most widely used and promising candidates for it.  The discovery of first experimental TSC, $Cu_x Bi_2 Se_3$ \cite{CuBi2Se3}, has generated considerable attention, as this system has been widely proposed as a promising platform for realizing topological superconductivity. In centrosymmetric superconductors, parity mixing of the pair pair potential may occur near surfaces where inversion symmetry is naturally broken, making them promising candidate for TSC \cite{centroAlphaPdBi2}.  A prime example is $\beta$-PdBi$_2$, a centrosymmetric superconductor that hosts topological surface states (TSSs) at the Fermi level in its normal state alongside fully gapped bulk superconductivity \cite{ARPESnewPdBi2,PdBi2ARPES}. It has been reported that centrosymmetric SnTaS$_2$ hosts topological nodal-line states and drumheadlike surface states which are quite visible near the Fermi level \cite{SntaS2,sntasanodal}. While p+ip systems are promising candidates for topological superconductivity, many candidate materials exhibiting p-wave superconductivity remain ambiguous \cite{Sr2RuO4arpes,PRLsr2Ruo4}. However, p-wave pairing can be realized on the interface of a heterostructure by the proximity between a strong topological insulator (TI) and an s-wave superconductor \cite{NbSe2Bi2Se3}.

Importantly, centrosymmetry and nonsymmorphicity are not mutually exclusive: a crystal space group may possess an inversion center while simultaneously containing nonsymmorphic symmetry operations. 
Thus the compound In$_2$Bi is a very fascinating example that combines these two aspects of crystal symmetry because its crystal geometry possesses an inversion center and $P6_3/mmc$ space group contains screw axes $(6_3)$ and glide planes. W.Kuang and et.al. \cite{In2Biadvancedmaterials} demonstrated that that its nonsymmorphic crystal symmetry, combined with SOC, generates nontrivial band topology which results in robust TSSs that coexist with bulk superconductivity and their first pricipal calculation exhibits multiband superconductivity. A similar case was also observed for centrosymmetric $\beta$-PdBi$_2$ \cite{FullygapPdBi2,multiphasePdBi2}. Magnetization and AC susceptibility measurements on In$_2$Bi single crystals have revealed anomalous electromagnetic responses (such as distinct signatures of critical fields like H$_{c2}$ and H$_{c3}$ ) tied directly to these protected surface states.
Our study shows that In$_2$Bi is type-II, single gap s-wave BCS superconductor in the dirty limit. Superfluid density $\rho_s(T)$ and electronic specific heat $\Delta C(T)$ consistily shows a fully-gapped superconductivity rather than a multiband superconductivity.

\section{\label{sec:level2}	    EXPERIMENTAL PROCEDURE}

The In$_2$Bi single crystal sample used in this study was prepared by the melt growth method with the multiple heat treatments. High purity elements, In balls (99.9999\%, Alfa Aesar) and Bi pieces (99.9999\%, Alfa Aesar) were weighed in proper stoichiometry ratio of 2 : 1, and sealed in an Argon purged quartz tube under high vacuum up to $10^{-6}$ mbar. In the first step, the ampule was loaded in box furnace and temperature was increased up to $500^\circ\mathrm{C}$ at a rate of $50^\circ\mathrm{C}\,\mathrm{h}^{-1}$ and held at this temperature for 48 hours for the better homogeneity of the sample and then naturally cool down to the room temperature. In the second step, ampule was heated at $200^\circ\mathrm{C}$ for 24 hours and after that the ampule was cooled down to $100^\circ\mathrm{C}$ at the rate of $5^\circ\mathrm{C}\,\mathrm{h}^{-1}$, and then slowly cooled down to $75^\circ\mathrm{C}$ at the rate of $1^\circ\mathrm{C}\,\mathrm{h}^{-1}$. To prevent oxidation and the decomposition of $\text{In}_2\text{Bi}$ into secondary phases such as $\text{In}_5\text{Bi}_3$ and $\text{InBi}$, the samples were stored in a high-purity argon-filled glove box.

The phase purity and crystal structure of the sample were examined using X-ray diffraction (XRD) with Cu-K$_\alpha$ radiation on a Rigaku HR-XRD diffractometer. Magnetization measurements were carried out using a Quantum Design Magnetic Property Measurement System (QD MPMS 3) on a rectangular shaped sample. Resistivity and specific heat measurements were measured using a Dynacool Quantum Design Physical Property Measurement System ( QD Dynacool PPMS). Resistivity was measured using a standard four-probe method on a rectangular sample, while specific heat measurements were performed using a thermal relaxation method, temperature down to 1.8 K.
Penetration depth measurements were carried out in a cryogen free magnetic system (CFMS) using a tunnel diode oscillator (TDO).

\section{\label{sec:level3}	 RESULTS AND DISCUSSIONS}

\subsection{\label{sec:level4} X-Ray Diffraction}
\begin{figure}[h!]
\centering
\includegraphics[width=0.5\textwidth]{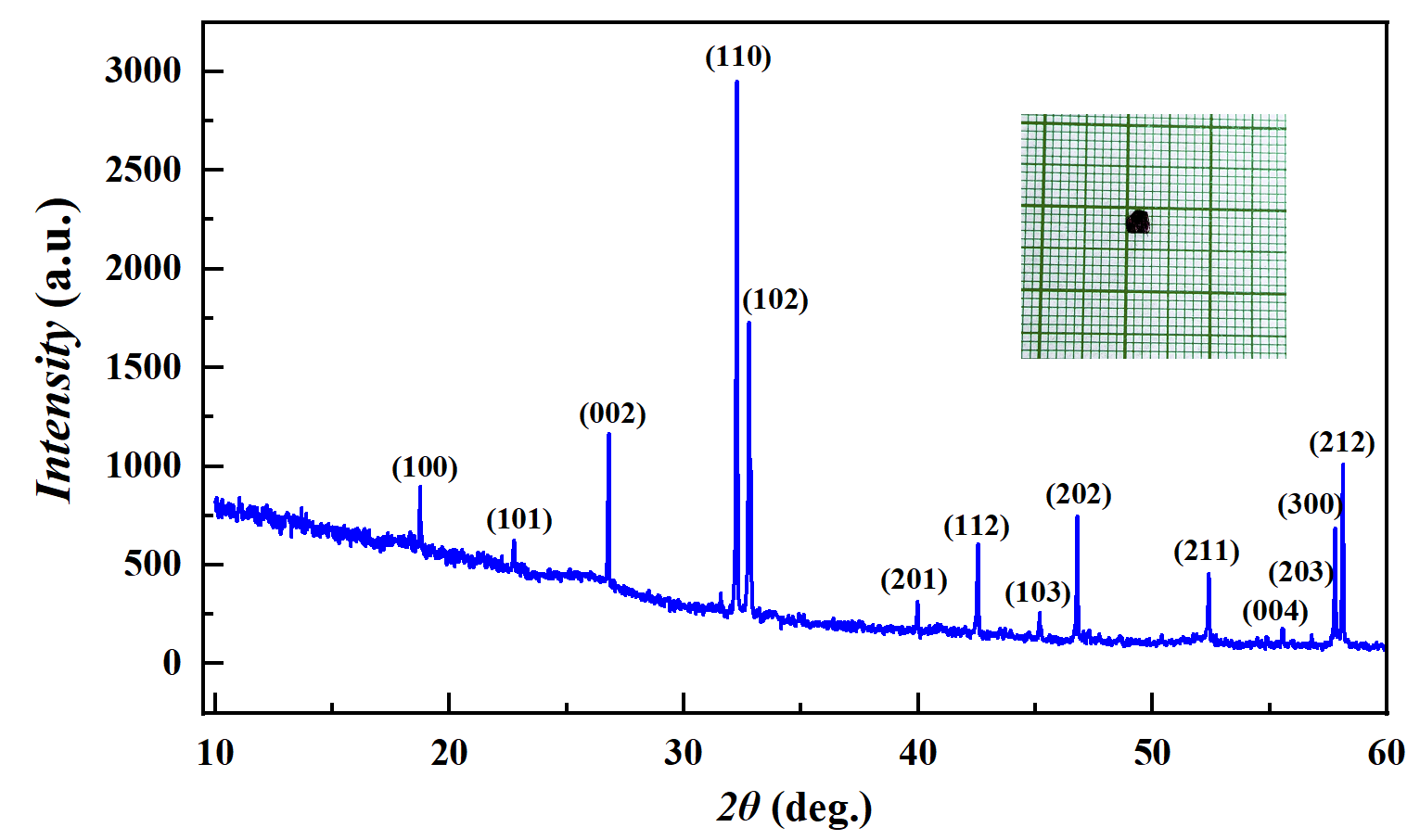}
\caption{shows X-ray diffraction (XRD) pattern for In$_2$Bi and inset shows the image of pallet with shiny metallic luster surface.}
\label{fig:XRDIn2Bi}
\end{figure}

Figure (\ref{fig:XRDIn2Bi}) shows the powder X-ray diffraction (XRD) pattern of polycrystalline In$_2$Bi. Prior to the measurement, the pellet was mechanically polished to achieve a shiny surface with a metallic luster, as shown in the inset. The diffraction pattern exhibits no impurity peaks, showing good agreement with previous reports\cite{In2Biadvancedmaterials}.

\subsection{\label{sec:level5} Resistivity}
\begin{figure*}[t]
    \centering
    \includegraphics[width=1\textwidth]{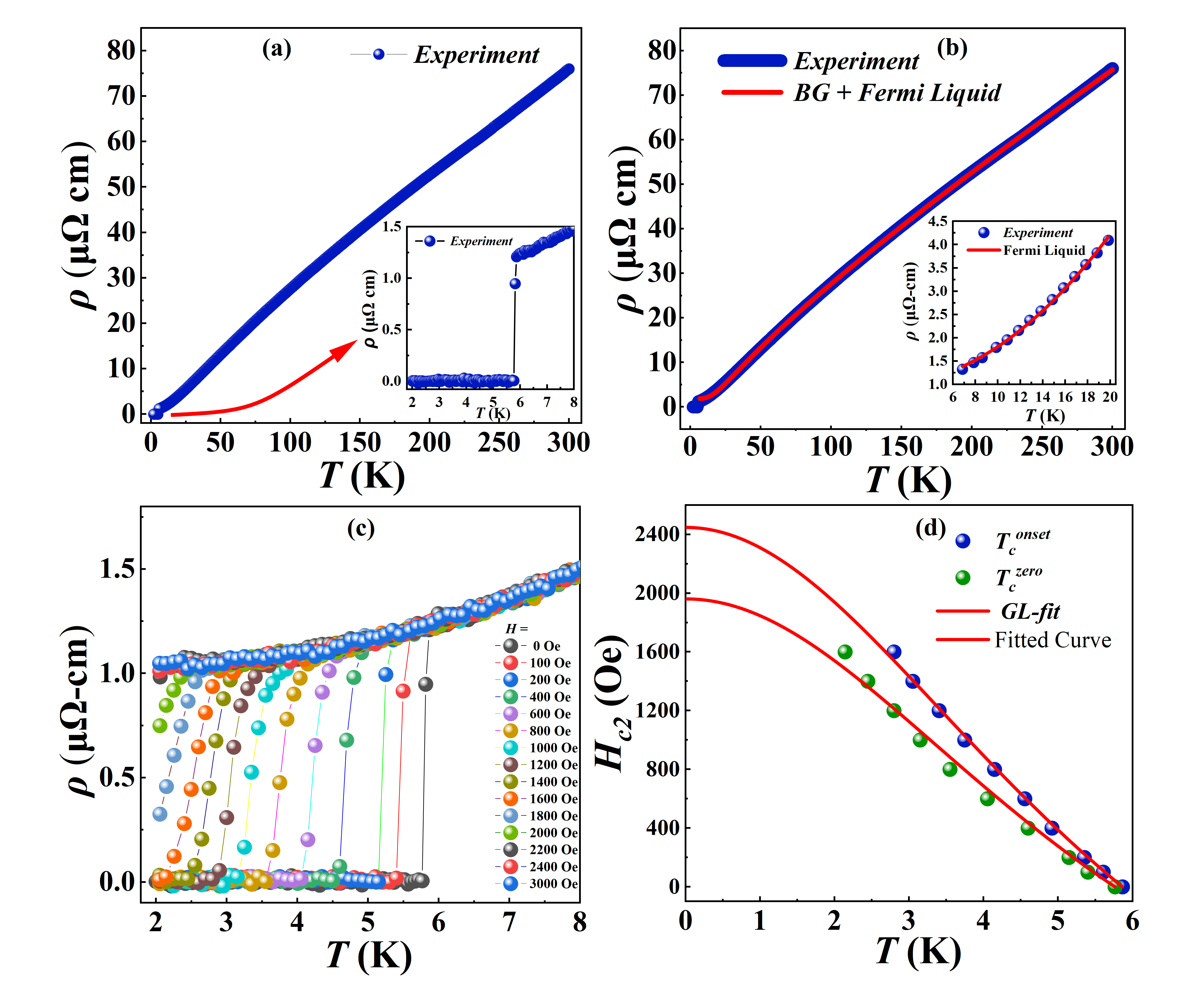}
    \caption{(a)shows the resistivity data in the temperature range 1.8 - 300 K. Lower inset is the close view near the superconducting transition. (b) shows the fitting using Bloch--Gr\"uneisen and Fermi liquid model in range of 7 - 300 K and upper inset shows the Fermi liquid fit in temperature range 7 - 20 K. (c) shows the magneto-resistance in temperature range 1.9 - 8 K under various magnetic fields.(d) Solid red curve shows the GL fit for the $T^{\text{onset}}_c$ and $T^{\text{zero}}_c$.}
    \label{fig:RMRall}
\end{figure*}

Figure \ref{fig:RMRall}(a) shows the resistivity data in temperature range 1.8 - 300 K, measured using standard four probe method in 10 mA applied current in zero magnetic field. The lower inset shows the superconducting transition at 5.86 K with the sharp superconducting gap $\Delta T (T_{c}^{onset}-T_{c}^{zero})$ = 0.1 K and the residual resistivity ratio (RRR), calculated as RRR = $\rho (300K)$/$\rho(7K)$, is 57.32. Both $\Delta T$ and RRR are reflecting the high quality of the sample. Above T$_c$, the sample shows typical metallic behavior. The lower inset of Fig.\ref{fig:RMRall}(b) shows the normal state resistivity data in temperature range of 7 - 20 K, which follow the Fermi liquid behavior for a typical metal and can be fit using the formula $\rho(T)$ = $\rho_0 + A_{e-e}{T^2}$, where $A_{e-e}$ corresponds to the electron-electron scattering effects and $\rho_0$ is residual resistivity.From the fitting, the residual resistivity $\rho_0 = 1.0097\,\mu\Omega\,\text{cm}$ and the electron--electron scattering coefficient $A_{e\text{--}e} = 8.03 \times 10^{-3}\,\mu\Omega\,\text{cm}\,\text{K}^{-2}$ are obtained. The close agreement with the values reported for single-crystalline In$_2$Bi suggests that grain-boundary scattering makes a negligible contribution to the resistivity of polycrystalline In$_2$Bi.
Upon increasing temperature, the resistivity deviates from the Fermi-liquid ($T^2$) behavior. 
To quantify the scattering mechanism we fitted the normal state resistivity in the temperature range of 7-300~K Fig.\ref{fig:RMRall}(b) using combination of Fermi liquid and Bloch--Gr\"uneisen (BG) model described as
\begin{equation}
\rho = \rho_0 + aT^2 + \rho(T)_{BG}.
\end{equation}   
where,
 \begin{equation}
\rho(T)_{BG} = 4R \left( \frac{T}{\Theta_{\rm D}} \right)^5 \int_0^{\Theta_{\rm D}/T} \frac{t^5}{(e^t - 1)(1 - e^{-t})} \, dt 
\label{eq:BG}
\end{equation}
where $\rho_{0}$ is the residual resistivity and second term describes the electron-phonon scattering, where $\theta{_D}$ is the Debye temperature and $R$ is coupling constant.
The BG fitting yields 
$\Theta_D = 95.195(69)\,\mathrm{K}$, $\rho_0 = 2.211(32)\,\mu\Omega\,\mathrm{cm}$, and $R = 101.54(70)\,\mu\Omega\,\mathrm{cm\,K^{-2}}$.
The obtained value of $\Theta_D$ is consistent with that estimated from specific-heat measurements.
[Eq.~\ref{eq:BG}], consistent with dominant electron--phonon scattering.
 The breakdown of the BG description at lower temperatures reflects a crossover between these scattering mechanisms prior to the onset of superconductivity.   
Figure~\ref{fig:RMRall}(c) displays the electrical resistivity measured under various magnetic fields applied parallel to the sample surface. The upper critical field values ($H_{c2}$) were determined from the resistivity-versus-temperature curves using two characteristic temperatures: $T_c^{\text{zero}}$, defined as the temperature where the resistivity drops to zero, and $T_c^{\text{onset}}$, corresponding to the temperature where the superconducting transition begins to deviate from the normal state.   

Instead of the Werthamer-Helfand-Hohenberg (WHH) model given in Eq.(3), we used the generalized Ginzburg-Landau (GL) model as described in Eq.(4), to calculate the upper critical field $H_{c2}$. Because the WHH model describes the orbital-limited $H_{c2}$ of type-II superconductors in the dirty limit, this fit significantly underestimates the experimental $H_{c2}$ values and fails to capture the upward curvature observed at lower temperatures.

\begin{equation}
H_{c2}(0) = -0.693 \, T_c \left[ \frac{dH_{c2}}{dT} \right]_{T=T_c}  
\label{eq:Hc2}
\end{equation}

\begin{equation} 
H_{c2}(T) = H_{c2}(0) \frac{1 - \left(\frac{T}{T_c}\right)^2}{1 + \left(\frac{T}{T_c}\right)^2}
\label{eq:Hc2}
\end{equation}

 The Figure ~\ref{fig:RMRall}(d) shows the phase diagrams of $H_{c2}(T)$ and solid red line represents the extrapolated GL model fitting for $T^{\text{onset}}_c$ and $T^{\text{zero}}_c$ obtained from  resistivity. As shown, the solid red line fits well with the GL-equation. The estimated values of $H_{c2}(0)$ are, $2448.08$ Oe for $T^{\text{onset}}_c$ and $1961.15$ Oe for $T^{\text{zero}}_c$, consistent with the previous report. The excellent agreement of the data with the GL model suggests the conventional superconductivity, indicating that orbital pairing mechanism is dominant in In$_2$Bi.
 The higher $H_{c2}(0)$ obtained from the onset criterion likely reflects superconductivity in regions with the highest local $H_{c2}$, while the zero-resistance criterion is influenced by vortex motion and phase-coherence effects.
 These values of $H_{c2}(0)$, while remaining significantly smaller than the Pauli limit $H_p = 1.84\,T_c \approx 10.79~\text{T}$ (i.e., $H_{c2}(0) \ll H_p$), indicate that Pauli paramagnetic effects are negligible and that $H_{c2}$ is orbital-limited.
 
In addition, we used $H_{c2}(0) = 2448.08$ Oe to calculate the Ginzburg-Landau coherence length, $\xi_{GL}$(0), using formula given below, 
\begin{equation}
H_{c2}(0) = \frac{\Phi_0}{2\pi \xi_{GL}(0)^2}
\end{equation}
Here, $\Phi_0 = 2.067 \times 10^{-15} \, \text{Wb}$ is the flux quantum. The estimated coherence length is $36.6$ nm, consistent with the previous report.

\subsection{\label{sec:level7} Magnetization}
\begin{figure*}[t]
    \centering
    \includegraphics[width=1\textwidth]{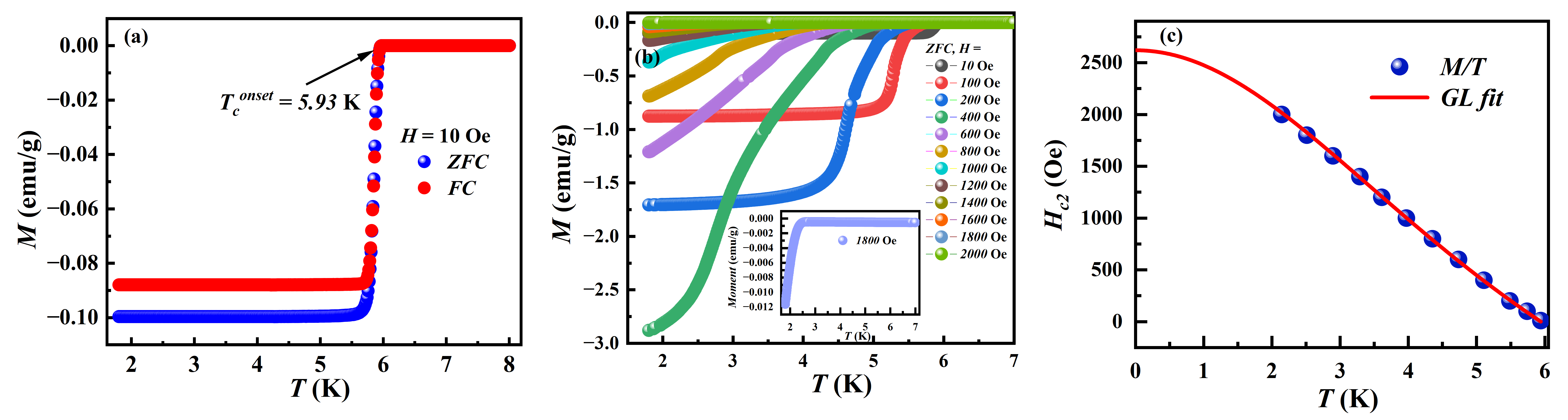}
   \caption{(a) Temperature dependence of magnetization measured in zero-field-cooled (ZFC) and field-cooled (FC) modes at $10\text{ Oe}$ between $1.8\text{ and }8\text{ K}$. (b) Magnetization curves under various magnetic fields up to $2000\text{ Oe}$, with the inset showing an enlarged view at $1800\text{ Oe}$. (c) Upper critical field $H_{c2}(T)$ fitted with the generalized GL model.}
    \label{fig:MTallin2bi}
\end{figure*}

Figure~\ref{fig:MTallin2bi}(a) shows the DC magnetization data for bulk In$_2$Bi in zero field cooled (ZFC) and field cooled (FC) mode under an applied magnetic field of 10 Oe. Both the FC and ZFC curves show the sharp drop in the magnetization at 5.93 K and the bifurcation in both the curves suggesting type-II superconductivity. Magnetization was also performed on the samle after 24 h air exposure. A kink was observed around 4.2 K indicating the presence of In$_5$Bi$_3$ as shown in the fig.(2) of supplementary. Figure~\ref{fig:MTallin2bi}(b) represents magnetization curves under an applied external field ranging from 0 to 2000 Oe and the inset shows the enlarged view at 1800 Oe field. Further to calculate the $H_{c2}(0)$, we noted down the points $T^{\text{onset}}_c$ above which the sample enters in the diamagnetic region. As shown in Fig.\ref{fig:MTallin2bi}(c), again generalized GL model was applied to calculate the $H_{c2}(0)$. The obtained value $H_{c2}(0) = 2619.3768$ Oe, is consistent with the previous values obtained from resistivity.

\begin{figure*}[t]
    \centering
    \includegraphics[width=1\textwidth]{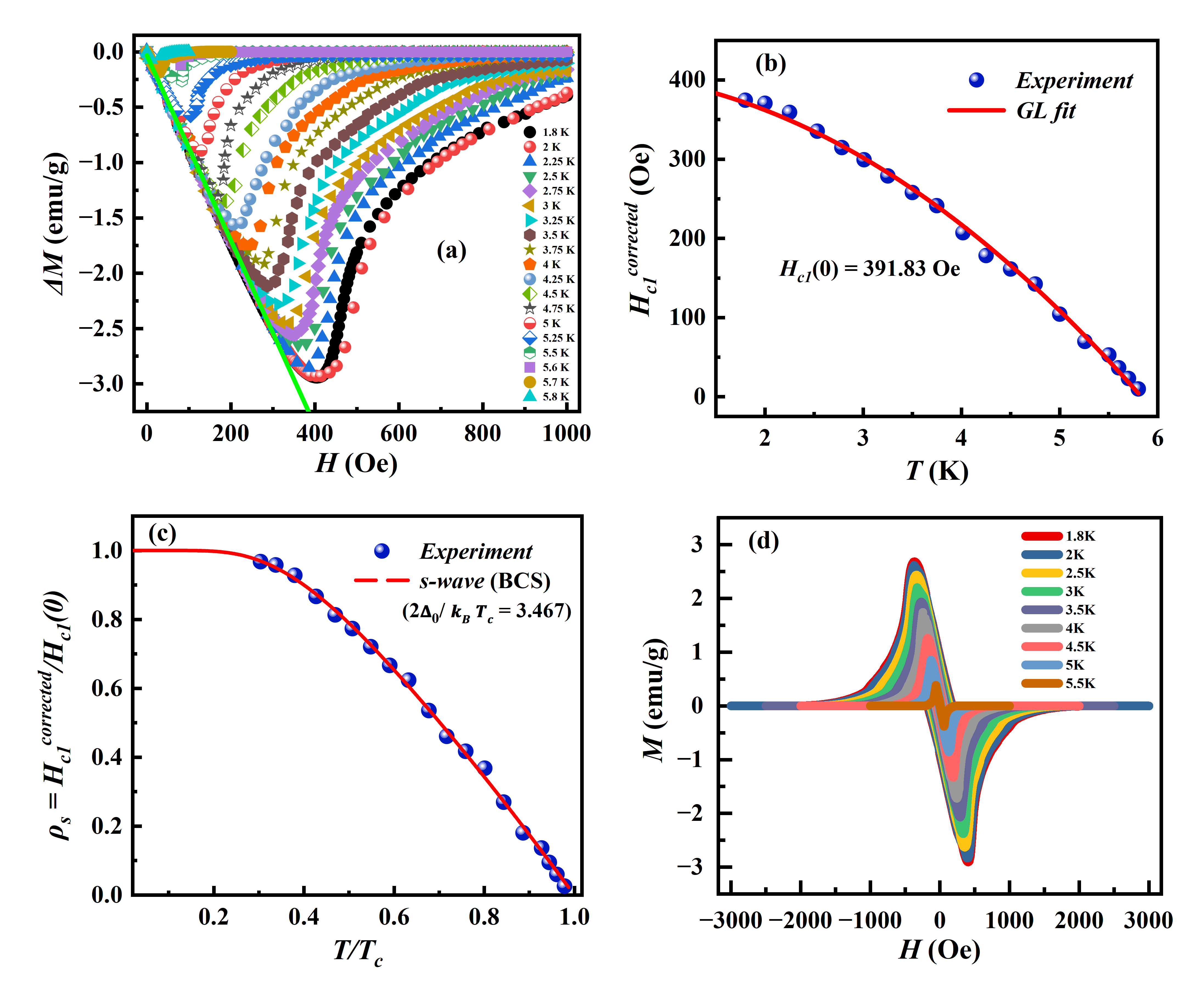}
    \caption{(a) M(H) isotherm curves in 0-1000 Oe magnetic field after the demagnetization correction. The solid green line is Meissner-line corresponding to the linear fitting.(b) shows the $H_{c1}(T)$ phase diagram and solid red line is the GL model fit.(c)  shows superfluid density ($\rho_s$(T)) calculated from $H_{c1}$ and solid red line is BCS s-wave fitting.(d) shows typical hysteresis loops for type-II superconductors at various temperatures.}
    \label{fig:MHin2biall}
\end{figure*}

Figure~\ref{fig:MHin2biall}(a) shows the field dependent magnetization curves M(H) at different temperatures. The lower critical field ($H_{c1}$) was determined from the initial isothermal M(H) curves at low fields where M(H) deviates from the initial Meissner state. In the perfect Meissner state, the sample behaves as a perfect diamagnet, yielding a linear relation between the applied field ($H_{app}$) and the magnetization. The experimental lower critical field ($Hc^{\text{expt}}_1$) was identified using the deviation method, where a linear fit was established in the low field regime. The deviation from the linearity, defined as $\Delta M = M_{expt} - M_{fit}$, was plotted against $H_{app}$, and onset of flux penetration ($Hc^{\text{expt}}_1$) was defined as the field where $\Delta M$ strictly deviates from zero.
Because the sample is a rectangular slab with dimensions (length l=3.260 mm, width w=2.683 mm and thickness t=0.367 mm), the internal magnetic field ($H_{int}$) differs from the $H_{app}$ due to the demagnetization effects. With the $H_{app}$ parallel to the length, the demagnetization factor ($N_{L}$) along the longitudinal axis was evaluated. For perfectly diamagnetic cuboid ($\chi = -1$ or M = -$H_{int}$), $H_{int} = H_{app}-N_{L}M$. The true thermodynamic lower critical field ($Hc^{\text{corrected}}_1$) was evaluated as $Hc^{\text{corrected}}_1 = Hc^{\text{expt}}_1/1+N_L$. To calculate $N_L$, we utilized the volume-averaged magnetometric demagnetization factor for a uniformly magnetized prism derived by Aharoni \cite{demagnetization}, yielding $N_L \approx 0.062$, model accounting for the non-uniform shielding currents typical of finite superconductors.
The lower critical field $H_{c1}$(0) was estimated using the Ginzburg Landau (GL) model,
\begin{equation}
   H_{c1}(T) = H_{c1}(0)\left[1 - \left(\frac{T}{T_c}\right)^2\right]
\end{equation}
as shown in Fig.\ref{fig:MHin2biall}(b). The obtained value of $H_{c1} = 391.83$ Oe. With $H_{c2}(0) = 2619.3768$ Oe, we obtained $\xi_{GL}$(0)=35.43 nm using Eq.(5). Penettration depth $\lambda_{GL}(0)$ was obtained using the relation,
\begin{equation}
H_{c1}= \frac{\Phi_0}{4\pi \lambda_{GL}^2}(ln\kappa + 0.5)
\end{equation}
where $\kappa = \lambda/\xi$. The estimated value of $\lambda_{GL}(0)$ and $\kappa$ are 101.9 nm and 2.88, consistent with the previous reports. All these values strongly indicate type-II superconductivity in In$_2$Bi.

The temperature dependence of $H_{c1}$ provides a powerful macroscopic window into the microscopic physics of the superconducting state. Because $H_{c1}$(T) scales inversely with the square of London penetration depth ($H_{c1} \propto \lambda^{-2}$) and serves as a direct probe of stiffness of superconducting condensate.
To extract the normalized superfluid density ($\rho_s (T)$), we utilized the London relation $\rho_s (T) = H_{c1}(T)/H_{c1}(0)$, where $H_{c1}(0)$ is the lower critical field at T = 0 K.

For an isotropic, single-gap $s$-wave superconductor, the temperature dependence of the normalized superfluid density $\rho_s(T)$ is described by the standard weak-coupling BCS theory as \cite{M.Tinkham}, 

\begin{equation}
\rho_s(T) = 1 - 2 \int_{\Delta(T)}^{\infty} \left( -\frac{\partial f}{\partial E} \right) \frac{E}{\sqrt{E^2 - \Delta(T)^2}} \, dE
\end{equation}

Where:
    $f(E) = \frac{1}{e^{E/k_B T} + 1}$ is the Fermi-Dirac distribution function. $E$ is the quasiparticle excitation energy.
$\Delta(T)$ is the temperature-dependent superconducting energy gap given as,
\begin{equation}
\Delta(T) = \Delta_0 \tanh \left[ 1.82 \left( 1.018 \left( \frac{1}{t} - 1 \right) \right)^{0.51} \right]
\end{equation},

with the reduced temperature $t = T/T_c$. For weak-coupling $s$-wave BCS superconductors, the zero-temperature gap scales universally with the critical transition temperature $T_c$, as $\Delta_0 = 1.764 \, k_B T_c$.
Figure~\ref{fig:MHin2biall}(c) represents the normalized $\rho_s(T)$ and solid red line is the single gap s-wave fit. The fitting yields $2\Delta_0/k_B T_c = 3.467$ which is little bit smaller than the weak coupling limit of BCS prediction $2\Delta_0/k_B T_c = 3.53$. A typical M(H) loops at various temperatures is shown in Fig.\ref{fig:MHin2biall}(d), indicate type-II superconductivity in In$_2$Bi.

\subsection{\label{sec:level6} Penetration Depth and Superfluid Density}
\begin{figure*}[t]
    \centering
    \includegraphics[width=1\textwidth]{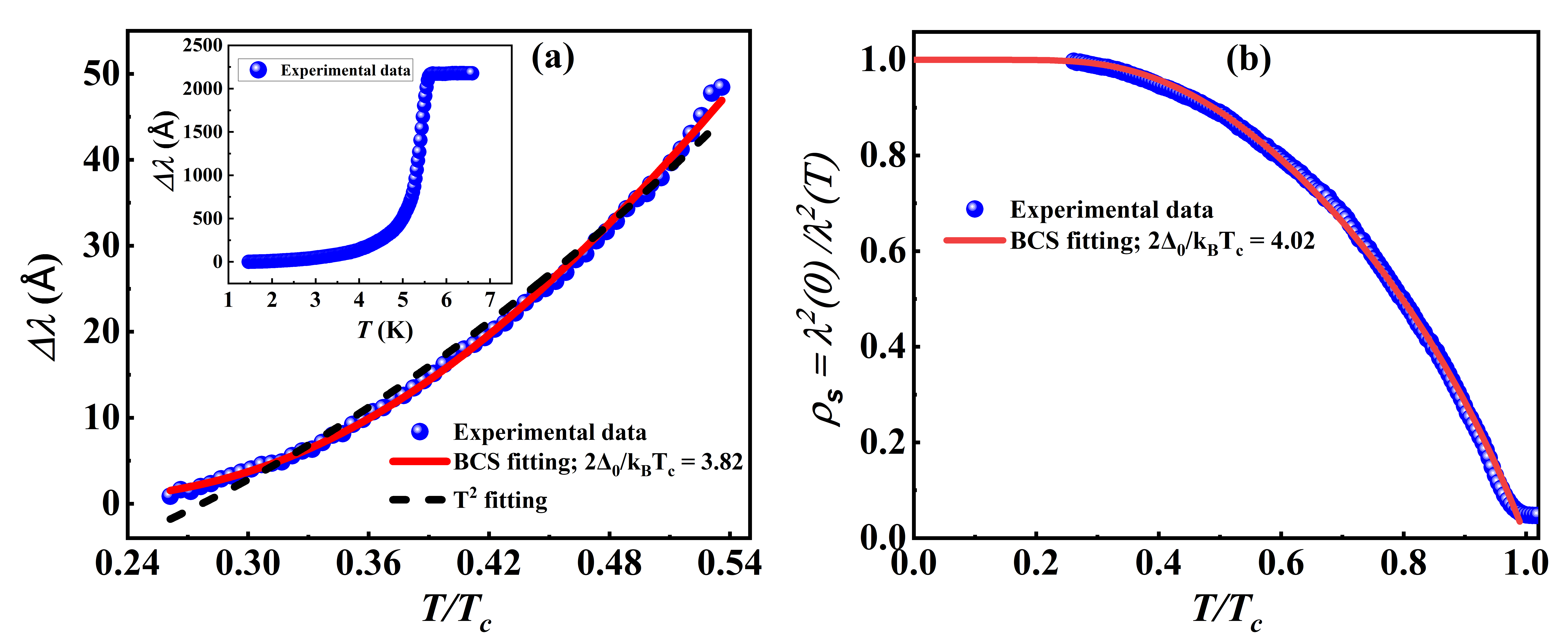}
    \caption{(a) shows low temperature $\Delta\lambda$ as a function of reduced temperature (T/T$_c$). Solid red line is the BCS fitting and dashed black represent $T^2$ fitting and inset shows the experimental $\Delta\lambda(T)$ down to temperature 1.6 K.(b) shows the superfluid density $\rho_s$ as a function of reduced temperature (T/T$_c$) and solid red line is the BCS fitting with a superconducting gap $2\Delta_0/k_B T_c = 4.02$.}
    \label{fig:SFDIn2Bi}
\end{figure*}

To gain more insight the superconducting region, London penetration depth is an important parameter to characterize superconducting pairing symmetries. The temperature dependence of London penetration depth, $\Delta \lambda(T)$, was measured using a TDO technique. This technique is highly sensitive to the small changes in the resonant frequency. The measured frequency shift, $\Delta F(T)$ was converted to change in the penetration depth as $\Delta F(T) = G \Delta \lambda(T)$, where G is a geometrical factor depends on the geometry of resonator coil of the LC-circuit. For our setup G = 2.27 $\AA$/Hz. The inset of Fig.\ref{fig:SFDIn2Bi}(a) shows the London penetration depth change $\Delta \lambda(T )$ of In$_2$Bi over a wider temperature range from 1.6 to 7 K.  A sharp increase in $\Delta\lambda(T )$ is observed near the superconducting transition temperature, $T_c \sim 5.68$ K, indicating a high-quality sample with a well defined superconducting transition as shown in the inset of Fig.\ref{fig:SFDIn2Bi}(a). 

The low-temperature data were first fitted using the conventional isotropic BCS model, which predicts an exponential temperature dependence of the penetration depth \cite{M.Tinkham},

\begin{equation}
\Delta\lambda(T)=\lambda(0)\sqrt{\frac{\pi\Delta_0}{2k_BT}}
\exp\left(-\frac{\Delta_0}{k_BT}\right),
\end{equation}

where $\Delta_0$ is the superconducting energy gap at zero temperature and $\lambda(0)$ is the zero-temperature penetration depth obtained from the upper critical field analysis. As shown in Fig.\ref{fig:SFDIn2Bi}(b), the BCS model fits the experimental data very well. The fit gives a superconducting gap ratio of $\frac{2\Delta_0}{k_BT_c}= 3.82$, which is higher than the weak-coupling BCS value of 3.53. This result suggests that In$_2$Bi is a strong coupled superconductor with a fully gapped $s$-wave superconducting gap.

To check whether the superconducting gap contains line nodes or point nodes, $\Delta \lambda(T)$ was also fitted using a power law temperature dependence for $T\textless T_c$,

\begin{equation}
\Delta\lambda(T)\propto T^n,
\end{equation}

For n = 1, which is expected for clean $d$-wave superconductors and n = 2, which is often associated with point nodes or impurity-scattered nodal superconductors. Thus, power-law models, Fig.\ref{fig:SFDIn2Bi}(a), fail to describe the low-temperature penetration depth as accurately as the exponential BCS model. These results provide strong evidence that In$_2$Bi is a fully gapped superconductor with conventional $s$-wave pairing symmetry. The larger value of $2\Delta_0/k_BT_c$ further indicates that superconductivity in In$_2$Bi lies in the strong-coupling regime.

To further examine the superconducting gap symmetry in In$_2$Bi, the normalized superfluid density, $\rho_s(T)$, was calculated from the temperature-dependent magnetic penetration depth using

\begin{equation}
\rho_s(T)=\left[\frac{\lambda(0)}{\lambda(T)}\right]^2,
\end{equation}

where $\lambda(T)=\lambda(0)+\Delta\lambda(T)$ and $\lambda(0)$ is the zero-temperature penetration depth obtained from the upper critical field analysis. The temperature dependence of the normalized superfluid density is shown in the Fig.\ref{fig:SFDIn2Bi}(b).

The experimental data were analyzed using the conventional single-gap isotropic $s$-wave BCS model, where the superfluid density is calculated using Eqs.~(8) and (9). In contrast, an isotropic $s$-wave fit with an adjustable gap parameter, $\frac{\Delta_0}{k_BT_c}$, reproduces the experimental data very well over the entire temperature range. The best fit yields a superconducting gap ratio of $\frac{2\Delta_0}{k_BT_c}=4.02$, which is larger than the weak-coupling BCS value of 3.53. Our results indicate that In$_2$Bi is a fully gapped $s$-wave superconductor in the strong-coupling regime in the dirty limit of BCS prediction.

\subsection{\label{sec:level4} Specific Heat}
\begin{figure*}[t!]
\centering
\includegraphics[width=1\textwidth]{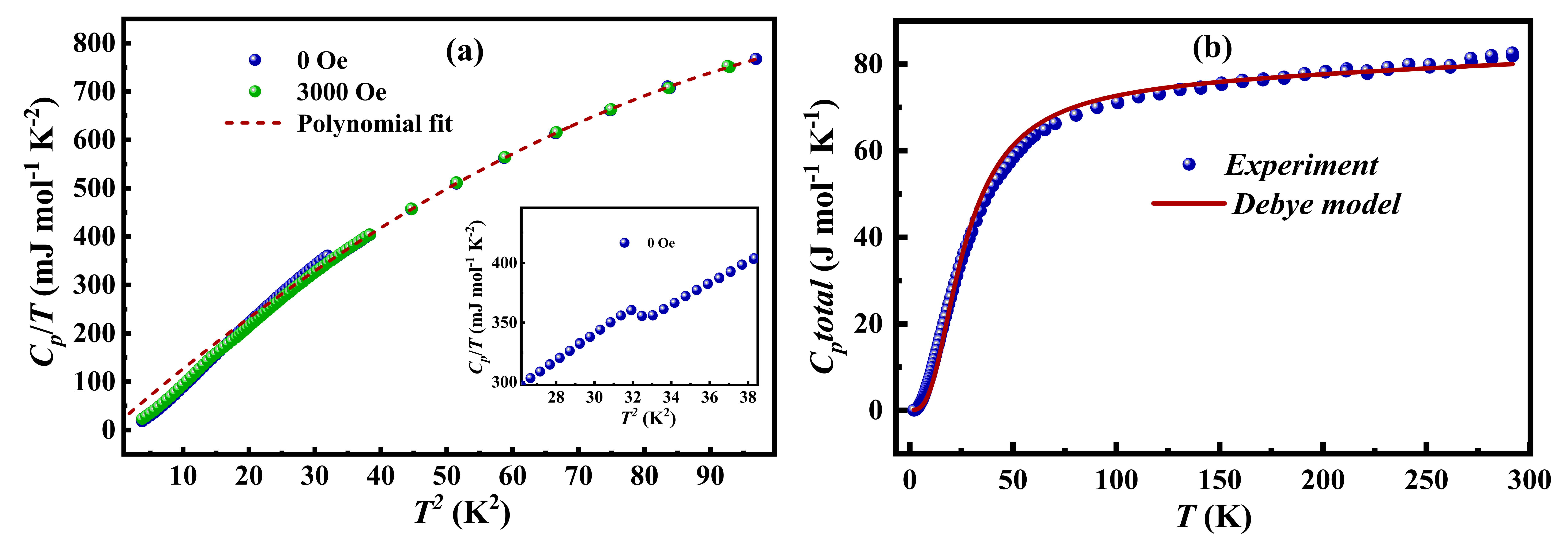}
\caption{(a) shows $C_p/T$ vs. $T^2$ of In$_2$Bi at magnetic fields $0$ and $3000$ Oe. The inset shows the close view of superconducting transition at 0 Oe field.(b) shows total specific heat C$_p$total(T) at 0 Oe field and fitted with Debye model.}
\label{fig:in2bi debye}
\end{figure*}

The Figure~\ref{fig:in2bi debye}(a) shows the low temperature specific heat  C$_p$/T vs. T$^2$ of In$_2$Bi at H = $0$ and $3000$ Oe magnetic fields Oe. In zero field, onset superconducting transition in $C_p$(T) is observed around $5.69$ K, a close view of the superconducting transition is shown in inset of Fig.\ref{fig:in2bi debye}(a). At 3000 Oe (more than H$_{c2}$(0)) magnetic field, superconductivity is fully suppressed. The normal-state specific heat $C_n$(T) is extracted from the H = 3000 Oe data above the superconducting transition temperature as shown in Fig.\ref{fig:in2bi debye}(a) by dashed red line. The normal state  C$_p$/T vs. T$^2$ at low temperatures shows non linear behavior. We therefore fitted $C_n$(T), given as, $C_n(T) = \gamma_n T + \beta T^3$ + $\delta T^5$, where the normal electronic contribution is denoted by $\gamma_n T$ and the phonon contribution by $\beta T^3$ and $\delta T^5$.
The fitting yields $\gamma_n = 10.62\, \text{mJ/mol K}^2$ and $\beta = 9.81\, \text{mJ/mol K}^4$ and $\delta = 0.04\, \text{mJ/mol K}^6$. From $\beta$, we derived the corresponding Debye temperature, $\theta_D = 106.11 $ K using the relation $\theta_D^3 = \frac{12 \pi^4 N R}{5 \beta}$, where R is the universal gas constant and N is the number of atoms per formula unit (here N = 6). These results are consistent with previous report \cite{In2Bimaterialstoday}. The DOS (Density of states) at Fermi energy N($E_F$) = $4.51$ states/(eV f.u.) was estimated using the relation $\gamma = \frac{\pi^2 k_B^2 N(E_F)}{3}$. This DOS N($E_F$) contains the enhancement factor due to the electron-phonon coupling constant  $\lambda_{\text{e-ph}}$, and can be related to the bare-band DOS at the Fermi level $N_0(E_F)$ by the expression $N(E_F) = N_0(E_F) \left( 1 + \lambda_{\text{e-ph}} \right)$. The value $\lambda_{\text{e-ph}}$ = 0.67 is consistent with previous report \cite{In2Bimaterialstoday} and obtained  from low temperature specific heat data using MacMillan’s formula described in equation (14). The Fermi energy, E$_f$ = 0.33 eV was estimated using the relation $E_f = \frac{\pi^2 k_B^2 N_A}{2\gamma}$, where N$_A$ is the Avogadro's number.
The total heat capacity ($C_p total$) in zero magnetic field as a function of temperature for $\text{In}_2\text{Bi}$ is shown in Fig. 5(b). In the normal state at high temperatures, the specific heat saturates to $C(T = 261\text{ K}) \sim 78.19\text{ J}\cdot\text{mol}^{-1}\cdot\text{K}^{-1}$, which is nearly close to the classical Dulong–Petit limit of $C = 3nR = 9R = 74.83\text{ J}\cdot\text{mol}^{-1}\cdot\text{K}^{-1}$ due to lattice vibrations, as expected \cite{PdTecsyadav}.
For more consistency, we also fitted the low-T normal state heat capacity data with a combination of the electronic and lattice contributions (Debye model of specific heat) using the expression given below,\cite{PdTecsyadav,C.Kittel} 
 \begin{equation}
C = \gamma_nT  + 9NR \left( \frac{T}{\Theta_{\rm D}} \right)^3 \int_0^{\Theta_{\rm D}/T} \frac{t^4 e^t}{(e^t - 1)^2} \, dt 
\end{equation}
The best possible fit gives the Sommerfeld coefficient $\gamma_n = 10.93 \, \text{mJ/mol K}^2$ and $\theta_D = 106.65 $ K. These values are consistent with the values obtained later.     

The relation of electronic specific heat $\Delta C(T) = C_p(T,H=0 Oe)-C_n(T)$ can be used to study the features of the superconducting phase transition in In$_2$Bi. The dimensionless specific-heat jump at transition temperature $T_c$ is $\frac{\Delta C}{\gamma_n T_c}$ = 1.61 which is greater than typical BCS weak-coupling value $(\approx 1.43)$, suggesting the strength of electron pairing and high quality of the sample.
The electron-phonon coupling constant, $\lambda_{\text{e-ph}}$ was calculated using inverted Macmillan's formula,  

\begin{equation}
\lambda_{\text{e-ph}} = \frac{\mu^* \ln \left( \frac{\theta_D}{1.45 T_c} \right) + 1.04}{\ln \left( \frac{\theta_D}{1.45 T_c} \right) (1 - 0.62 \mu^*)-1.04} 
\end{equation}
Taking $\theta_D$ = 106.11 K, $T_c$ = 5.69 K and $\mu^* \approx 0.13$, the value of $\lambda_{\text{e-ph}}$ = 1.05 falls at the boundary of, and conventionally just within, the strong-coupling regime..

\begin{figure}[t!]
\centering
\includegraphics[width=0.5\textwidth]{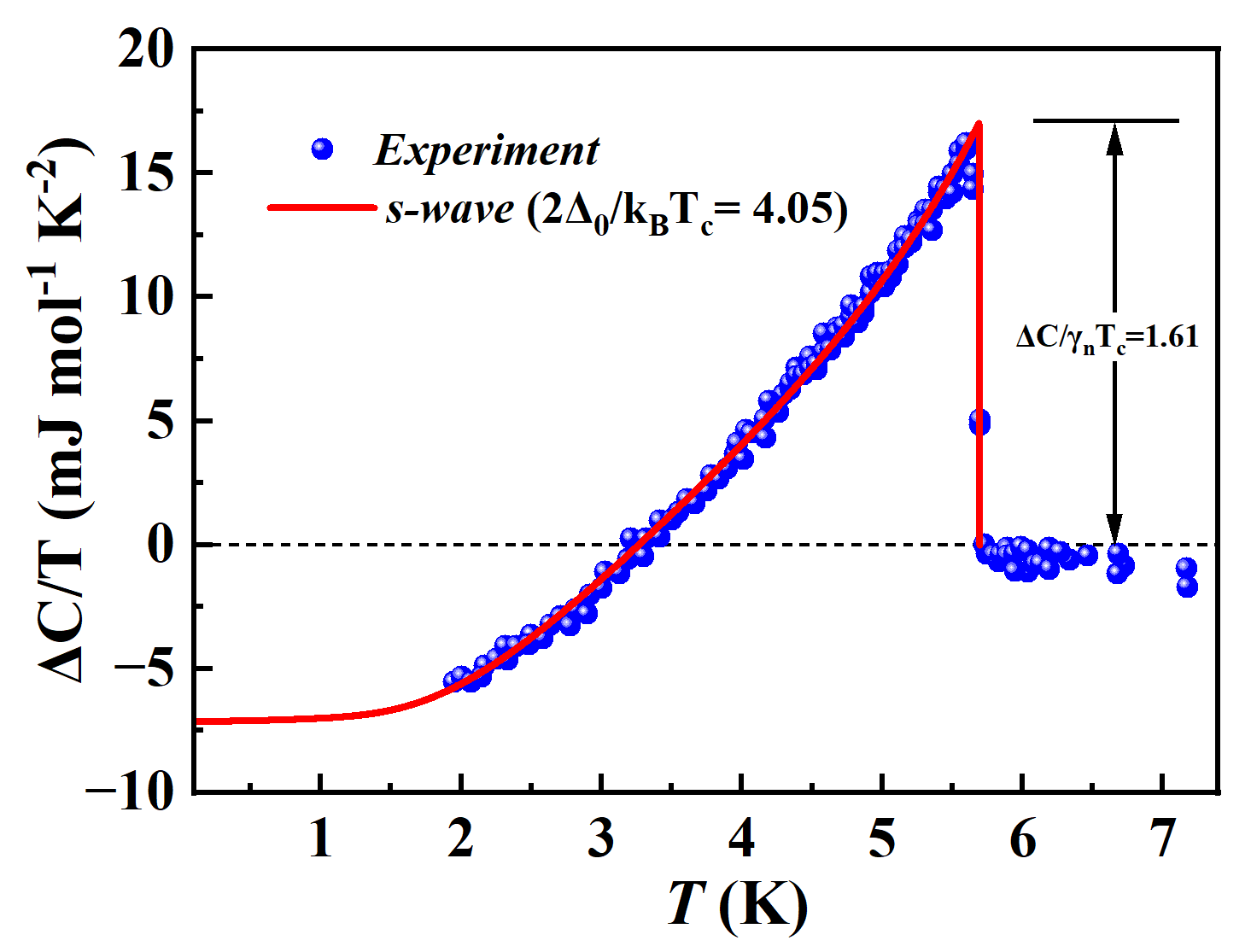}
\caption{shows temperature dependence of normal state specific heat divided by temperature, $\Delta C/T$ vs.T. The solid red line is the s-wave BCS fitting with $2\Delta_0/k_B T_c =4.05$.}
\label{fig:CelIn2Bi}
\end{figure}

Based on the normal state specific heat $C_n$ obtained above, the electronic specific heat under zero field $C_{\text{el}}(T)$ was estimated as shown in Fig.6, where $\Delta C$ represents the difference between specific heats under $0$ and $3000\text{ Oe}$: as
$\frac{\Delta C}{T} = \frac{C(T)}{T} - \frac{C_n(T)}{T} = \frac{C_{\text{el}}(T)}{T}-\gamma$, \cite{PRLSrPt3P}.

Electronic specific heat was fitted to a single-gap s-wave, in order to investigate the superconducting order parameter of In$_2$Bi using the subsequent formula \cite{alphamodelpadm}, 
\begin{equation}
\begin{aligned}
S_{es}(T) = - \frac{6\gamma_n}{\pi^2} \frac{\Delta(0)}{k_B}\int_0^\infty dy \left[ f \ln f + (1-f) \ln(1-f) \right],
\end{aligned}
\end{equation}
\begin{equation}
C_{\mathrm{es}} = T\frac{dS}{dT}    
\end{equation}
where $f=\frac{1}{e^{\beta E} + 1}$ and $\beta = 1/k_B T$, whereas the integration variable $y=\epsilon/\Delta_0$. Bogoliubov quasiparticle energy is given by $E = \sqrt{\epsilon^2 + \Delta^2(T)}$, and $\epsilon$ is electron energy with respect to the Fermi energy. $\Delta(T)$ is the gap function for an isotropic s-wave as described in Eq.(9) and $\delta(T) = \frac{\Delta(T)}{\Delta_0}$ is the normalized BCS gap at the reduced temperature $t=T/T_c$ as tabulated by Mühlschlegel \cite{alphamodel}. By employing so called $\alpha$-model \cite{alphamodelpadm}, where $\alpha=\Delta_0/k_B T_c$ is the only adjustable parameter which provides a means of scaling the BCS gap as $\Delta(T) = \left(\frac{\alpha}{\alpha_{BCS}}\right) \Delta_{\mathrm{0}}$, where $\alpha_{BCS}=1.764$.

The Fig.(\ref{fig:CelIn2Bi}) shows that experimental data is well fitted with $\alpha=\Delta_0/k_B T_c\sim 2.025$. The obtained value $2\Delta_0/k_B T_c = 4.05$ is slightly larger than the weak-coupling BCS value $2\Delta_0/k_B T_c = 3.53$. The strong coupling s-wave superconductivity in In$_2$Bi with $2\Delta_0/k_B T_c = 4.05$ is in good agreement with some Pd/Pb-Bi base alloys \cite{PdBi2Cp, PRBTlBi2}, A15 compounds \cite{Ti3SbRPS, Ti3SbRchap}, Chevrel phases \cite{CheveralCp}, and pyrochlore osmates \cite{pyrochloreoxideCp}.

The gap values obtained from the specific heat and superfluid density measurements are in good agreement with those extracted from the lower critical field, $H_{c1}$, and penetration depth, $\lambda(T)$, measurements, as listed in Table I. Although these values are slightly higher, they consistently support a single-gap, strong-coupling superconducting scenario in In$_2$Bi.
\begin{table}[t]
\centering
\caption{Comparison of the superconducting gap values obtained from different experimental probes.}
\label{tab:gap_comparison}
\large
\renewcommand{\arraystretch}{1.3}
\setlength{\tabcolsep}{9pt}
\begin{tabular}{lcccc}
\hline\hline
Probe & $H_{c1}(T)$ & $\lambda(T)$ & $\rho_s(T)$ & $C_{\mathrm{el}}(T)$ \\
\hline
$2\Delta_0/k_{\mathrm B}T_c$
& 3.467 & 3.82 & 4.02 & 4.05 \\
\hline\hline
\end{tabular}
\end{table}
    
\section{\label{sec:level6} Conclusion}
In summary, we have investigated the structural, normal-state, and superconducting properties of polycrystalline $\mathrm{In_2Bi}$ through complementary electrical resistivity, magnetization, magnetic penetration-depth, and specific-heat measurements. A sharp superconducting transition is consistently observed around $T_c\approx6$~K, with $T_c=5.86$~K from resistivity, $5.93$~K from magnetization, $5.69$~K from specific heat, and $5.68$~K from the penetration-depth measurements, demonstrating the reproducibility of the superconducting transition across different experimental probes. The high residual-resistivity ratio of $57.32$ further indicates the good quality of the investigated sample.

The field-dependent resistivity and magnetization measurements establish the type-II nature of superconductivity in $\mathrm{In_2Bi}$. The upper critical field extrapolated from the generalized Ginzburg--Landau description lies in the range $H_{c2}(0)\approx2.0$--$2.6$~kOe, depending on the criterion used, while the magnetization analysis gives $H_{c2}(0)=2619$~Oe. The corresponding coherence length is approximately $35$--$37$~nm. The lower critical field is found to be $H_{c1}(0)=391.8$~Oe after accounting for the demagnetization correction, yielding $\lambda_{\mathrm{GL}}(0)\approx101.9$~nm and $\kappa\approx2.88$. These parameters place $\mathrm{In_2Bi}$ firmly in the type-II superconducting regime. Moreover, the experimentally determined $H_{c2}(0)$ is far below the Pauli limiting field, supporting an orbital-limited upper critical field.

The temperature dependence of the superconducting condensate provides further evidence for a fully opened superconducting gap. The normalized superfluid density obtained from $H_{c1}(T)$ is well described by an isotropic single-gap $s$-wave BCS model, with $2\Delta(0)/k_BT_c=3.467$. Although this value is close to the weak-coupling BCS value of 3.53, the independent TDO penetration-depth measurements yield $2\Delta_0/k_BT_c=4.02$, while the low-temperature exponential behavior gives a value of approximately 3.82. Importantly, the exponential temperature dependence of $\Delta\lambda(T)$ provides a substantially better description of the low-temperature data than power-law forms, arguing against line or point nodes and multiband superconductivity as described by W.Kuang and et.al \cite{In2Biadvancedmaterials}.

The specific-heat measurements independently support this picture. The Sommerfeld coefficient and Debye temperature are found to be approximately $\gamma_n=10.6$--$10.9$~mJ~mol$^{-1}$~K$^{-2}$ and $\Theta_D\approx106$~K, respectively. The normalized specific-heat jump, $\Delta C/\gamma_nT_c=1.61$, exceeds the weak-coupling BCS value of 1.43, while the electron--phonon coupling constant estimated from the inverted McMillan relation is $\lambda_{e-ph}\approx1.05$. Furthermore, the electronic specific heat is satisfactorily described by an isotropic single-gap model with $2\Delta_0/k_BT_c=4.05$. The agreement between the specific-heat and penetration-depth analyses, both yielding gap magnitudes larger than the weak-coupling BCS expectation, provides thermodynamic as well as electromagnetic evidence for a fully gapped superconducting state with appreciable electron--phonon coupling.

Taken together, the results establish $\mathrm{In_2Bi}$ as a fully gapped, predominantly isotropic $s$-wave type-II superconductor with appreciable electron--phonon coupling. The consistency among transport, magnetic, penetration-depth, and thermodynamic measurements indicates that the bulk superconducting state is well described within a conventional phonon-mediated pairing framework, while the relatively enhanced gap ratio and specific-heat jump point toward coupling beyond the weak-coupling BCS limit. Given the nonsymmorphic $P6_3/mmc$ crystal symmetry and the previously reported topological surface states in $\mathrm{In_2Bi}$ \cite{In2Biadvancedmaterials}, the coexistence of these states with a fully gapped bulk superconducting condensate makes this material particularly interesting for further investigations of the interplay between crystal symmetry, surface electronic states, and superconductivity. However, direct experimental identification of a topological superconducting surface state is beyond the scope of the present measurements and requires surface-sensitive or phase-sensitive probes.

\section{\label{sec:level7} ACKNOWLEDGMENTS}
A. Gangwar is thankful to the Indian Institute of Technology (BHU), Varanasi – 221005, for providing a teaching assistantship. The author also gratefully acknowledges the Central Instrument Facility (CIF), IIT (BHU) for providing the necessary characterization/analytical facilities.

\bibliography{bibfile}

\end{document}